\documentclass{article}
\usepackage{amssymb}
\usepackage{amsmath}
\usepackage{amsfonts}
\usepackage{sw20ssur}
\usepackage{cite}
\usepackage{afterpage}

\input{tcilatex}
\begin{document}

\title{{\Large A game theoretical perspective on the quantum probabilities
associated with a GHZ state}}
\author{Azhar Iqbal and Derek Abbott \\
{\small School of Electrical \& Electronic Engineering, the University of
Adelaide,}\\
{\small South Australia 5005, Australia.}}
\maketitle

\begin{abstract}
In the standard approach to quantum games, players' moves are local unitary
transformations on an entangled state that is subsequently measured.
Players' payoffs are then obtained as expected values of the entries in the
payoff matrix of the classical game on a set of quantum probabilities
obtained from the quantum measurement. In this paper, we approach quantum
games from a diametrically opposite perspective. We consider a classical
three-player symmetric game along with a known expression for a set of
quantum probabilities relevant to a tripartite Einstein-Podolsky-Rosen (EPR)
experiment that depends on three players' directional choices in the
experiment. We define the players' moves as their directional choices in an
EPR setting and then express their payoff relations in the resulting quantum
game in terms of their directional choices, the entries of the payoff
matrix, and the quantum probability distribution relevant to the tripartite
EPR experiment.
\end{abstract}

\section{ Introduction}

In the standard scheme \cite{EWL, EW} of a quantized version of a
non-cooperative game \cite{GoogleScholar}, the players share an entangled
state, their strategic moves are local unitary transformations on the state,
and the quantum measurement \cite{Peres} generates the players' payoffs. The
resulting players' payoffs in the quantum game can be understood as the
expected values of the entries in the payoff matrix of the (classical) game 
\cite{Binmore, Rasmusen, Osborne} arising from a set of quantum
probabilities \cite{Peres}. The key concerns in determining the players'
payoffs relations in the quantum game are a) What are the players' moves in
the quantum game? b) Which set of quantum \ probabilities is obtained by
quantum measurement? and c) How the players' strategic moves are related to
the set of quantum probabilities?

This brings us to question whether the unitary transformations are really
necessary in the setup of a quantum game. A proposed scheme \cite%
{IqbalWeigert, Iqbal, Iqbal1, Chappell} for playing a quantum game in which
players' strategic moves are not unitary transformations uses the setting of
an Einstein-Podolsky-Rosen (EPR) experiment \cite{Peres, Bell1, Bell2,Bell3,
Aspect, CHSH}. Two players are located in spacelike-separated regions and
share a singlet state. In a run of the experiment, each player decides one
out of the two available directions and a quantum measurement is performed.
This leads to obtaining a (normalized) set of quantum probabilities along
with a listing of the directional choices the players make in each run of
the experiment. As the players' directional choices determine the quantum
probability distribution, the setting can be used to develop a quantum
version of a two-player game. A multipartite EPR experiment would then be
required for a multiplayer quantum game.

In this paper, we consider a classical three-player symmetric game, along
with a reported expression for a quantum probability distribution, which is
relevant to the three-partite EPR experiment. We then define players'
directional choices in the experiment as their strategic moves and express
players' payoff relations in the quantum game in terms of the three
directional choices and the entries of the payoff matrix.

This paper thus provides a game-theoretic perspective on the peculiarity of
quantum probabilities. The first perspective along game-theoretical lines on
quantum probabilities that are associated to the GHZ state \cite{Peres} was
provided by Vaidman in Ref. \cite{Vaidman}. Vaidman proposed a set of rules
defining a game that can only be won by a team of three players when they
share a GHZ state. The present paper extends Vaidman's perspective by
considering Nash equilibria in the set of symmetric games played by a team
of three players in a non-cooperative game setting. Vaidman presented his
game without invoking Hilbert space as is the case in the present paper.

\section{Three-player games with mixed-strategies}

Consider a three-player (noncooperative) game in which the players Alice $%
(A) $, Bob $(B)$, and Chris $(C)$ make their strategic moves simultaneously.
The players are assumed located at distance and are unable to communicate to
one another. They, however, can communicate to a referee who organizes the
game and ensures that the rules of the game are obeyed. Each player has to
decide between two choices, called the \textit{pure strategies}, and in
repeated version of the game they can also play the \textit{mixed strategies}%
. Their payoff relations are made public by the referee at the start of the
game. The payoff relations depend on the game matrix, the players' pure
strategies, and the probability distribution on pure strategies.

To be specific, we assume that the player $A$'s pure strategies are $S_{1},$ 
$S_{2}$; the player $B$'s pure strategies are $S_{1}^{\prime },$ $%
S_{2}^{\prime }$; and the player $C$'s pure strategies are $S_{1}^{\prime
\prime },$ $S_{2}^{\prime \prime }$. Also, the game is defined by the
following pure-strategy payoff relations \cite{IqbalToor}%
\begin{equation}
\begin{array}{l}
\Pi _{A,B,C}(S_{1},S_{1}^{\prime },S_{1}^{\prime \prime })=\alpha _{1},\beta
_{1},\gamma _{1}; \\ 
\Pi _{A,B,C}(S_{2},S_{1}^{\prime },S_{1}^{\prime \prime })=\alpha _{2},\beta
_{2},\gamma _{2}; \\ 
\Pi _{A,B,C}(S_{1},S_{2}^{\prime },S_{1}^{\prime \prime })=\alpha _{3},\beta
_{3},\gamma _{3}; \\ 
\Pi _{A,B,C}(S_{1},S_{1}^{\prime },S_{2}^{\prime \prime })=\alpha _{4},\beta
_{4},\gamma _{4};%
\end{array}%
\begin{array}{l}
\Pi _{A,B,C}(S_{1},S_{2}^{\prime },S_{2}^{\prime \prime })=\alpha _{5},\beta
_{5},\gamma _{5}; \\ 
\Pi _{A,B,C}(S_{2},S_{1}^{\prime },S_{2}^{\prime \prime })=\alpha _{6},\beta
_{6},\gamma _{6}; \\ 
\Pi _{A,B,C}(S_{2},S_{2}^{\prime },S_{1}^{\prime \prime })=\alpha _{7},\beta
_{7},\gamma _{7}; \\ 
\Pi _{A,B,C}(S_{2},S_{2}^{\prime },S_{2}^{\prime \prime })=\alpha _{8},\beta
_{8},\gamma _{8}.%
\end{array}
\label{payoffs_constants_definitions}
\end{equation}

For example, $\Pi _{A,B,C}(S_{1},S_{2}^{\prime },S_{1}^{\prime \prime
})=\alpha _{3},$ $\beta _{3},$ $\gamma _{3}$ states that the players $A$, $B$%
, and $C$ obtain the payoffs $\alpha _{3},$ $\beta _{3},$ and $\gamma _{3}$,
respectively, when they play the pure strategies $S_{1},$ $S_{2}^{\prime },$
and $S_{1}^{\prime \prime }$, respectively.

In a repeated version of this game, a player can choose between his/her two
pure strategies with some probability, which defines his/her mixed-strategy.
We specify a mixed-strategy by $x,$ $y,$ $z\in \lbrack 0,1]$ for players $A$%
, $B$, and $C$, respectively. These are the probabilities with which the
players $A$, $B$, and $C$ play the pure strategies $S_{1},$ $S_{1}^{\prime
}, $ and $S_{1}^{\prime \prime }$, respectively. They, then, play the pure
strategies $S_{2},$ $S_{2}^{\prime },$ and $S_{2}^{\prime \prime }$ with
probabilities $(1-x),$ $(1-y),$ and $(1-z)$, respectively, and the
mixed-strategy payoff relations, therefore, read%
\begin{equation}
\begin{array}{l}
\Pi _{A,B,C}(x,y,z)=xyz\Pi _{A,B,C}(S_{1},S_{1}^{\prime },S_{1}^{\prime
\prime })+x(1-y)z\Pi _{A,B,C}(S_{1},S_{2}^{\prime },S_{1}^{\prime \prime })+
\\ 
xy(1-z)\Pi _{A,B,C}(S_{1},S_{1}^{\prime },S_{2}^{\prime \prime
})+x(1-y)(1-z)\Pi _{A,B,C}(S_{1},S_{2}^{\prime },S_{2}^{\prime \prime })+ \\ 
(1-x)yz\Pi _{A,B,C}(S_{2},S_{1}^{\prime },S_{1}^{\prime \prime
})+(1-x)(1-y)z\Pi _{A,B,C}(S_{2},S_{2}^{\prime },S_{1}^{\prime \prime })+ \\ 
(1-x)y(1-z)\Pi _{A,B,C}(S_{2},S_{1}^{\prime },S_{2}^{\prime \prime
})+(1-x)(1-y)(1-z)\Pi _{A,B,C}(S_{2},S_{2}^{\prime },S_{2}^{\prime \prime }),%
\end{array}
\label{3coin_mixed_strategy_payoffs}
\end{equation}%
that can also be written as%
\begin{equation}
\begin{array}{l}
\Pi _{A,B,C}(x,y,z)=\dsum\limits_{i,j,k=1,2}\Pr_{c}(S_{i},S_{j}^{\prime
},S_{k}^{\prime \prime })\Pi _{A,B,C}(S_{i},S_{j}^{\prime },S_{k}^{\prime
\prime }),%
\end{array}%
\end{equation}%
where $\Pr_{c}(S_{i},S_{j}^{\prime },S_{k}^{\prime \prime })$ are the
classical factorizable probabilities and for instance, $%
\Pr_{c}(S_{1},S_{1}^{\prime },S_{2}^{\prime \prime })=xy(1-z)$ and $%
\Pr_{c}(S_{2},S_{2}^{\prime },S_{1}^{\prime \prime })=(1-x)(1-y)z.$

\subsection{Symmetric three-player games}

Symmetric three-player games are defined by the condition that a player's
payoff is decided by his/her strategic move and not by his/her identity.
Mathematically, this is expressed by the conditions%
\begin{equation}
\Pi _{A}(x,y,z)=\Pi _{A}(x,z,y)=\Pi _{B}(y,x,z)=\Pi _{B}(z,x,y)=\Pi
_{C}(y,z,x)=\Pi _{C}(z,y,x),  \label{constraints_for_symmetric_game}
\end{equation}%
i.e. the player $A$'s payoff when s/he plays $x$ remains the same either
when player $B$ plays $y$ whereas player $C$ plays $y$ or when player $B$
plays $x$ whereas player $C$ play $x$. The payoff relations (\ref%
{3coin_mixed_strategy_payoffs}) satisfy the conditions (\ref%
{constraints_for_symmetric_game}) when \cite{IqbalToor}%
\begin{equation}
\begin{array}{c}
\begin{array}{cccc}
\beta _{1}=\alpha _{1}, & \beta _{2}=\alpha _{3}, & \beta _{3}=\alpha _{2},
& \beta _{4}=\alpha _{3}, \\ 
\beta _{5}=\alpha _{6}, & \beta _{6}=\alpha _{5}, & \beta _{7}=\alpha _{6},
& \beta _{8}=\alpha _{8}, \\ 
\gamma _{1}=\alpha _{1}, & \gamma _{2}=\alpha _{3}, & \gamma _{3}=\alpha
_{3}, & \gamma _{4}=\alpha _{2}, \\ 
\gamma _{5}=\alpha _{6}, & \gamma _{6}=\alpha _{6}, & \gamma _{7}=\alpha
_{5}, & \gamma _{8}=\alpha _{8},%
\end{array}
\\ 
\begin{array}{cc}
\alpha _{6}=\alpha _{7}, & \alpha _{3}=\alpha _{4}.%
\end{array}%
\end{array}%
\end{equation}%
A symmetric three-player game can, therefore, be defined by only six
constants $\alpha _{1},$ $\alpha _{2},$ $\alpha _{3},$ $\alpha _{5},$ $%
\alpha _{6},$ and $\alpha _{8}$. In the rest of this paper we will define
these six constants to be $\alpha ,$ $\beta ,$ $\delta ,$ $\epsilon ,$ $%
\theta ,$ and $\omega ,$ where $\alpha _{1}=\alpha ,$ $\alpha _{2}=\beta ,$ $%
\alpha _{3}=\delta ,$ $\alpha _{5}=\epsilon ,$ $\alpha _{6}=\theta ,$ and $%
\alpha _{8}=\omega $. The pure-strategy payoff relations (\ref%
{payoffs_constants_definitions}) in this symmetric game are then
re-expressed as%
\begin{equation}
\begin{array}{l}
\Pi _{A,B,C}(S_{1},S_{1}^{\prime },S_{1}^{\prime \prime })=\alpha ,\alpha
,\alpha ; \\ 
\Pi _{A,B,C}(S_{2},S_{1}^{\prime },S_{1}^{\prime \prime })=\beta ,\delta
,\delta ; \\ 
\Pi _{A,B,C}(S_{1},S_{2}^{\prime },S_{1}^{\prime \prime })=\delta ,\beta
,\delta ; \\ 
\Pi _{A,B,C}(S_{1},S_{1}^{\prime },S_{2}^{\prime \prime })=\delta ,\delta
,\beta ;%
\end{array}%
\begin{array}{l}
\Pi _{A,B,C}(S_{1},S_{2}^{\prime },S_{2}^{\prime \prime })=\epsilon ,\theta
,\theta ; \\ 
\Pi _{A,B,C}(S_{2},S_{1}^{\prime },S_{2}^{\prime \prime })=\theta ,\epsilon
,\theta ; \\ 
\Pi _{A,B,C}(S_{2},S_{2}^{\prime },S_{1}^{\prime \prime })=\theta ,\theta
,\epsilon ; \\ 
\Pi _{A,B,C}(S_{2},S_{2}^{\prime },S_{2}^{\prime \prime })=\omega ,\omega
,\omega .%
\end{array}
\label{symmetric_3player_game_definition}
\end{equation}%
The mixed-strategy payoff relations in Eq. (\ref%
{3coin_mixed_strategy_payoffs}) then take the form

\begin{equation}
\begin{array}{l}
\Pi _{A,B,C}(x,y,z)=xyz(\alpha ,\alpha ,\alpha )+x(1-y)z(\delta ,\beta
,\delta )+xy(1-z)(\delta ,\delta ,\beta )+ \\ 
x(1-y)(1-z)(\epsilon ,\theta ,\theta )+(1-x)yz(\beta ,\delta ,\delta
)+(1-x)(1-y)z(\theta ,\theta ,\epsilon )+ \\ 
(1-x)y(1-z)(\theta ,\epsilon ,\theta )+(1-x)(1-y)(1-z)(\omega ,\omega
,\omega ).%
\end{array}
\label{Payoffs_factorizable}
\end{equation}

\section{Quantum probability distribution for a GHZ state}

Consider the GHZ state

\begin{equation}
\left\vert \psi \right\rangle =(\left\vert 0\right\rangle _{1}\left\vert
0\right\rangle _{2}\left\vert 0\right\rangle _{3}+\left\vert 1\right\rangle
_{1}\left\vert 1\right\rangle _{2}\left\vert 1\right\rangle _{3})/\sqrt{2},
\label{GHZ state}
\end{equation}%
that is shared among three the three players, where $\left\vert
i\right\rangle _{j}$ is the $i$-th state of the $j$-th qubit and the setting
of the generalized EPR experiments. Each player measures the dichotomic
observable $\vec{n}.\vec{\sigma}$ where $\vec{n}=\vec{a},\vec{b},\vec{c}$
and $\vec{\sigma}$ is a vector the components of which are standard Pauli
matrices. The family of observables $\vec{n}.\vec{\sigma}$ covers all
possible dichotomic observables for a qubit system \cite{Peres}.

Kaszlikowski and \.{Z}ukowski \cite{Kaszlikowski} show that the probability
of obtaining the result $m=\pm 1$ for the player $A$, when s/he plays the
strategy $\vec{a}$, the result $l=\pm 1$ for the player $B$, when s/he plays
the strategy $\vec{b}$ and the result $k=\pm 1$ for the player $C$, when
s/he plays the strategy $\vec{c}$ is given by

\begin{equation}
\Pr_{QM}(m,l,k;\vec{a},\vec{b},\vec{c})=\frac{1}{8}\left[
1+mla_{3}b_{3}+mka_{3}c_{3}+lkb_{3}c_{3}+mlk%
\sum_{r,p,s=1}^{3}M_{rps}a_{r}b_{p}c_{s}\right] ,
\label{KaszlikowskiEquation}
\end{equation}%
where $a_{r}$, $b_{p}$, $c_{s}$ are components of vectors $\vec{a},\vec{b},%
\vec{c}$ and where nonzero elements of the tensor $M_{rps}$ are $M_{111}=1,$ 
$M_{122}=-1,$ $M_{212}=-1,$ $M_{221}=-1$. In view of this, the only terms in
the product $a_{r}b_{p}c_{s}$ that contribute towards the probability $%
\Pr_{QM}(m,l,k;\vec{a},\vec{b},\vec{c})$ are $a_{1}b_{1}c_{1},$ $%
a_{1}b_{2}c_{2},$ $a_{2}b_{1}c_{2},$ and $a_{2}b_{2}c_{1}$. Eq. (\ref%
{KaszlikowskiEquation}) can therefore be written as

\begin{equation}
\Pr_{QM}(m,l,k;\vec{a},\vec{b},\vec{c})=\frac{1}{8}\left[
1+mla_{3}b_{3}+mka_{3}c_{3}+lkb_{3}c_{3}+mlk(a_{1}b_{1}c_{1}-a_{1}b_{2}c_{2}-a_{2}b_{1}c_{2}-a_{2}b_{2}c_{1})%
\right] .  \label{QProbabilities}
\end{equation}%
Note that Eq. (\ref{KaszlikowskiEquation}) gives a quantum probability
distribution without reference to the undelying Hilbert space, unitary
transformations, or quantum measurement.

We consider playing a three-player quantum game in which the players Alice,
Bob, and Chris (henceforth, labelled as player $A$, player $B$, and player $%
C $) moves consist of choosing the directions $\vec{a},\vec{b},$ and $\vec{c}
$, respectively. The players's payoff relations are then expressed in terms
of the quantum probability distribution given in Eq. (\ref%
{KaszlikowskiEquation}).

\subsection{Players sharing a GHZ state and when choosing a direction is a
player's move}

Let $\vec{a}=\vec{a}(a_{1},a_{2},a_{3}),$ $\vec{b}=\vec{b}%
(b_{1},b_{2},b_{3}),$ $\vec{c}=\vec{c}(c_{1},c_{2},c_{3})$ be the players'
directional choices that we consider as their strategies. Denoting the
quantum probabilities by $\Pr_{{\small Q}}$, the set of quantum
probabilities can be obtained from Eq. (\ref{QProbabilities}) as follows%
\begin{eqnarray}
\Pr_{{\small Q}}(S_{1},S_{1}^{\prime },S_{1}^{\prime \prime }) &=&\Pr_{%
{\small Q}}[(\vec{a},m=+1),(\vec{b},l=+1),(\vec{c},k=+1)]  \notag \\
&=&\frac{1}{8}\left[ 1+a_{3}b_{3}+a_{3}c_{3}+b_{3}c_{3}+\Delta \right] ; 
\notag \\
\Pr_{{\small Q}}(S_{1},S_{2}^{\prime },S_{1}^{\prime \prime }) &=&\Pr_{%
{\small Q}}[(\vec{a},m=+1),(\vec{b},l=-1),(\vec{c},k=+1)]  \notag \\
&=&\frac{1}{8}\left[ 1-a_{3}b_{3}+a_{3}c_{3}-b_{3}c_{3}-\Delta \right] ; 
\notag \\
\Pr_{{\small Q}}(S_{1},S_{1}^{\prime },S_{2}^{\prime \prime }) &=&\Pr_{%
{\small Q}}[(\vec{a},m=+1),(\vec{b},l=+1),(\vec{c},k=-1)]  \notag \\
&=&\frac{1}{8}\left[ 1+a_{3}b_{3}-a_{3}c_{3}-b_{3}c_{3}-\Delta \right] ; 
\notag \\
\Pr_{{\small Q}}(S_{1},S_{2}^{\prime },S_{2}^{\prime \prime }) &=&\Pr_{%
{\small Q}}[(\vec{a},m=+1),(\vec{b},l=-1),(\vec{c},k=-1)]  \notag \\
&=&\frac{1}{8}\left[ 1-a_{3}b_{3}-a_{3}c_{3}+b_{3}c_{3}+\Delta \right] ; 
\notag \\
\Pr_{{\small Q}}(S_{2},S_{1}^{\prime },S_{1}^{\prime \prime }) &=&\Pr_{%
{\small Q}}[(\vec{a},m=-1),(\vec{b},l=+1),(\vec{c},k=+1)]  \notag \\
&=&\frac{1}{8}\left[ 1-a_{3}b_{3}-a_{3}c_{3}+b_{3}c_{3}-\Delta \right] ; 
\notag \\
\Pr_{{\small Q}}(S_{2},S_{2}^{\prime },S_{1}^{\prime \prime }) &=&\Pr_{%
{\small Q}}[(\vec{a},m=-1),(\vec{b},l=-1),(\vec{c},k=+1)]  \notag \\
&=&\frac{1}{8}\left[ 1+a_{3}b_{3}-a_{3}c_{3}-b_{3}c_{3}+\Delta \right] ; 
\notag \\
\Pr_{{\small Q}}(S_{2},S_{1}^{\prime },S_{2}^{\prime \prime }) &=&\Pr_{%
{\small Q}}[(\vec{a},m=-1),(\vec{b},l=+1),(\vec{c},k=-1)]  \notag \\
&=&\frac{1}{8}\left[ 1-a_{3}b_{3}+a_{3}c_{3}-b_{3}c_{3}+\Delta \right] ; 
\notag \\
\Pr_{{\small Q}}(S_{2},S_{2}^{\prime },S_{2}^{\prime \prime }) &=&\Pr_{%
{\small Q}}[(\vec{a},m=-1),(\vec{b},l=-1),(\vec{c},k=-1)]  \notag \\
&=&\frac{1}{8}\left[ 1+a_{3}b_{3}+a_{3}c_{3}+b_{3}c_{3}-\Delta \right] ;
\label{QProbabs}
\end{eqnarray}%
where $\Delta
=a_{1}b_{1}c_{1}-a_{1}b_{2}c_{2}-a_{2}b_{1}c_{2}-a_{2}b_{2}c_{1}.$ We define
players $A$'s, $B$'s, $C$'s payoff relations in the quantum game as follows%
\begin{equation}
\begin{array}{l}
\Pi _{A,B,C}(\vec{a},\vec{b},\vec{c})=\sum_{i,j,k=1}^{2}%
\Pr_{Q}(S_{i},S_{j}^{\prime },S_{k}^{\prime \prime })\Pi
_{A,B,C}(S_{i},S_{j}^{\prime },S_{k}^{\prime \prime }),%
\end{array}
\label{Payoffs}
\end{equation}%
i.e. these are obtained as the expectation of payoff entries (\ref%
{symmetric_3player_game_definition}) on the set of quantum probabilities (%
\ref{QProbabs}). For the symmetric game defined in Eq. (\ref%
{symmetric_3player_game_definition}), the payoffs to the players $A,$ $B,$
and $C,$ given in (\ref{Payoffs}), can then be expanded as follows:

\begin{equation}
\begin{array}{l}
\Pi _{A,B,C}(\vec{a},\vec{b},\vec{c})= \\ 
\frac{1}{8}\{\left[ 1+a_{3}b_{3}+a_{3}c_{3}+b_{3}c_{3}+\Delta \right]
(\alpha ,\alpha ,\alpha )+\left[ 1-a_{3}b_{3}+a_{3}c_{3}-b_{3}c_{3}-\Delta %
\right] (\delta ,\beta ,\delta )+ \\ 
\left[ 1+a_{3}b_{3}-a_{3}c_{3}-b_{3}c_{3}-\Delta \right] (\delta ,\delta
,\beta )+\left[ 1-a_{3}b_{3}-a_{3}c_{3}+b_{3}c_{3}+\Delta \right] (\epsilon
,\theta ,\theta )+ \\ 
\left[ 1-a_{3}b_{3}-a_{3}c_{3}+b_{3}c_{3}-\Delta \right] (\beta ,\delta
,\delta )+\left[ 1+a_{3}b_{3}-a_{3}c_{3}-b_{3}c_{3}+\Delta \right] (\theta
,\theta ,\epsilon )+ \\ 
\left[ 1-a_{3}b_{3}+a_{3}c_{3}-b_{3}c_{3}+\Delta \right] (\theta ,\epsilon
,\theta )+\left[ 1+a_{3}b_{3}+a_{3}c_{3}+b_{3}c_{3}-\Delta \right] (\omega
,\omega ,\omega )\}.%
\end{array}
\label{Payoffs_ABC}
\end{equation}%
Let $a_{3}=b_{3}=c_{3}=0$ i.e. when the players' unit vectors are confined
to the X-Y plane, the payoff relations (\ref{Payoffs_ABC}) can be written as

\begin{equation}
\begin{array}{l}
\Pi _{A,B,C}(\vec{a},\vec{b},\vec{c})= \\ 
\frac{1}{8}\{(1+\Delta )(\alpha ,\alpha ,\alpha )+(1-\Delta )(\delta ,\beta
,\delta )+(1-\Delta )(\delta ,\delta ,\beta )+(1+\Delta )(\epsilon ,\theta
,\theta )+ \\ 
(1-\Delta )(\beta ,\delta ,\delta )+(1+\Delta )(\theta ,\theta ,\epsilon
)+(1+\Delta )(\theta ,\epsilon ,\theta )+(1-\Delta )(\omega ,\omega ,\omega
)\}.%
\end{array}
\label{ConfinedDirections}
\end{equation}%
It is apparent from above that the resulting payoff relations (\ref%
{ConfinedDirections}) in the quantum game cannot be put into a form that is
same as for the classical mixed-strategy game i.e. Eq. (\ref%
{Payoffs_factorizable}). This raises the question whether there exist
constraints that can be placed on the players' directional choices, i.e. the
unit vectors $\vec{a},$ $\vec{b},$ and $\vec{c},$ such that the payoff
relations (\ref{Payoffs_ABC}) in the quantum game are reduced to the
players' payoffs in the classical game allowing mixed strategies (\ref%
{Payoffs_factorizable}). In order to find an answer to this we set

\begin{equation}
\Pi _{A,B,C}(\vec{a},\vec{b},\vec{c})=\Pi _{A,B,C}(x,y,z),
\end{equation}%
and equate the right sides of Eqs. (\ref{Payoffs_ABC}, \ref%
{Payoffs_factorizable}) i.e.

\begin{eqnarray}
\frac{1}{8}\left[ 1+a_{3}b_{3}+a_{3}c_{3}+b_{3}c_{3}+\Delta \right] &=&xyz,
\label{Eq_1} \\
\frac{1}{8}\left[ 1-a_{3}b_{3}+a_{3}c_{3}-b_{3}c_{3}-\Delta \right]
&=&x(1-y)z,  \label{Eq_2} \\
\frac{1}{8}\left[ 1+a_{3}b_{3}-a_{3}c_{3}-b_{3}c_{3}-\Delta \right]
&=&xy(1-z),  \label{Eq_3} \\
\frac{1}{8}\left[ 1-a_{3}b_{3}-a_{3}c_{3}+b_{3}c_{3}+\Delta \right]
&=&x(1-y)(1-z),  \label{Eq_4} \\
\frac{1}{8}\left[ 1-a_{3}b_{3}-a_{3}c_{3}+b_{3}c_{3}-\Delta \right]
&=&(1-x)yz,  \label{Eq_5} \\
\frac{1}{8}\left[ 1+a_{3}b_{3}-a_{3}c_{3}-b_{3}c_{3}+\Delta \right]
&=&(1-x)(1-y)z,  \label{Eq_6} \\
\frac{1}{8}\left[ 1-a_{3}b_{3}+a_{3}c_{3}-b_{3}c_{3}+\Delta \right]
&=&(1-x)y(1-z),  \label{Eq_7} \\
\frac{1}{8}\left[ 1+a_{3}b_{3}+a_{3}c_{3}+b_{3}c_{3}-\Delta \right]
&=&(1-x)(1-y(1-z).  \label{Eq_8}
\end{eqnarray}%
Now, by adding Eqs. (\ref{Eq_1}) and (\ref{Eq_2}) we obtain

\begin{equation}
\frac{1}{4}(1+a_{3}c_{3})=xz,  \label{Eq_1.1}
\end{equation}%
adding Eqs. (\ref{Eq_1}) and (\ref{Eq_3}) gives

\begin{equation}
\frac{1}{4}(1+a_{3}b_{3})=xy,  \label{Eq_9}
\end{equation}%
adding Eqs. (\ref{Eq_1}) and (\ref{Eq_5}) gives

\begin{equation}
\frac{1}{4}(1+b_{3}c_{3})=yz,  \label{Eq_10}
\end{equation}%
adding Eqs. (\ref{Eq_3}) and (\ref{Eq_4}) gives

\begin{equation}
\frac{1}{4}(1-a_{3}c_{3})=x(1-z).  \label{Eq_11}
\end{equation}

Now, we add Eqs. (\ref{Eq_1.1}) and (\ref{Eq_11}) to obtain $x=\frac{1}{2}.$
Adding Eqs. (\ref{Eq_7}) and (\ref{Eq_8}) gives

\begin{equation}
\frac{1}{4}(1+a_{3}c_{3})=(1-x)(1-z),  \label{Eq_12}
\end{equation}%
and substitution from Eq. (\ref{Eq_1.1}) and $x=\frac{1}{2}$ gives $z=\frac{1%
}{2}.$ Similarly, adding Eqs. (\ref{Eq_2}) and (\ref{Eq_4}) gives

\begin{equation}
\frac{1}{4}(1-a_{3}b_{3})=x(1-y),  \label{Eq_13}
\end{equation}%
and adding Eqs. (\ref{Eq_6}) and (\ref{Eq_8}) gives

\begin{equation}
\frac{1}{4}(1+a_{3}b_{3})=(1-x)(1-y).  \label{Eq_14}
\end{equation}%
By adding Eqs. (\ref{Eq_13}) and (\ref{Eq_14}) we obtain $y=\frac{1}{2}$ and
thus $(x,y,z)=(\frac{1}{2},\frac{1}{2},\frac{1}{2})$ is obtained as the
solution of the Eqs. (\ref{Eq_1}) to (\ref{Eq_8}).

Therefore, the mixed-strategy payoff relations (\ref{Payoffs_factorizable})
can be recovered from the payoffs relations (\ref{Payoffs_ABC}) for the
quantum game only for the special case when $(x,y,z)=(\frac{1}{2},\frac{1}{2}%
,\frac{1}{2})$. This is because the quantum probability distribution for the
GHZ state, from which the payoff relations (\ref{Payoffs_ABC}) are
constructed, are inherently non-factorizable. In the research area of
quantum games, recovering the mixed strategy classical payoff relations from
the payoff relations for a quantum game is quite often considered an
essential requirement. When the underlying quantum probabilities in a
quantum game are obtained from the GHZ state, this requirement is not
satisfied except for a very special case, i.e. $(x,y,z)=(\frac{1}{2},\frac{1%
}{2},\frac{1}{2})$.

Considering the payoff relations (\ref{Payoffs_ABC}) in the quantum game, a
Nash equilibrium (NE) is a directional triple $(\vec{a}^{\ast },\vec{b}%
^{\ast },\vec{c}^{\ast })$ that satisfies the following constraints:

\begin{eqnarray}
\Pi _{A}(\vec{a}^{\ast },\vec{b}^{\ast },\vec{c}^{\ast })-\Pi _{A}(\vec{a},%
\vec{b}^{\ast },\vec{c}^{\ast }) &\geq &0,  \notag \\
\Pi _{B}(\vec{a}^{\ast },\vec{b}^{\ast },\vec{c}^{\ast })-\Pi _{B}(\vec{a}%
^{\ast },\vec{b},\vec{c}^{\ast }) &\geq &0,  \notag \\
\Pi _{C}(\vec{a}^{\ast },\vec{b}^{\ast },\vec{c}^{\ast })-\Pi _{C}(\vec{a}%
^{\ast },\vec{b}^{\ast },\vec{c}) &\geq &0,
\end{eqnarray}%
for all $\vec{a},$ $\vec{b},$ and $\vec{c}$. For the symmetric game, these
Nash inequalities take the form

\begin{eqnarray}
\Pi _{A}(\vec{a}^{\ast },\vec{b}^{\ast },\vec{c}^{\ast })-\Pi _{A}(\vec{a},%
\vec{b}^{\ast },\vec{c}^{\ast }) &=&\frac{1}{8}\left[ (a_{3}^{\ast
}-a_{3})\Delta _{1}\gamma _{1}+\gamma _{2}(a_{1}^{\ast }-a_{1})\Delta
_{2}-\gamma _{2}(a_{2}^{\ast }-a_{2})\Delta _{3}\right] \geq 0,  \notag \\
\Pi _{A}(\vec{a}^{\ast },\vec{b}^{\ast },\vec{c}^{\ast })-\Pi _{A}(\vec{a}%
^{\ast },\vec{b},\vec{c}^{\ast }) &=&\frac{1}{8}\left[ (b_{3}^{\ast
}-b_{3})\Delta _{1}^{\prime }\gamma _{1}-\gamma _{2}(b_{2}^{\ast
}-b_{2})\Delta _{2}^{\prime }+\gamma _{2}(b_{1}^{\ast }-b_{1})\Delta
_{3}^{\prime }\right] \geq 0,  \notag \\
\Pi _{A}(\vec{a}^{\ast },\vec{b}^{\ast },\vec{c}^{\ast })-\Pi _{A}(\vec{a}%
^{\ast },\vec{b}^{\ast },\vec{c}) &=&\frac{1}{8}\left[ (c_{3}^{\ast
}-c_{3})\Delta _{1}^{\prime \prime }\gamma _{1}-\gamma _{2}(c_{2}^{\ast
}-c_{2})\Delta _{2}^{\prime \prime }+\gamma _{2}(c_{1}^{\ast }-c_{1})\Delta
_{3}^{\prime \prime }\right] \geq 0,  \notag \\
&&  \label{NE_ABC}
\end{eqnarray}%
where

\begin{equation}
\gamma _{1}=\alpha -\beta -\epsilon +\omega \text{ and }\gamma _{2}=\alpha
-2\delta -\beta +\epsilon +2\theta -\omega ,
\end{equation}%
and

\begin{eqnarray}
\Delta _{1} &=&b_{3}+c_{3},\text{ }\Delta _{2}=b_{1}c_{1}-b_{2}c_{2},\text{ }%
\Delta _{3}=b_{1}c_{2}+b_{2}c_{1},  \notag \\
\Delta _{1}^{\prime } &=&a_{3}+c_{3},\text{ }\Delta _{2}^{\prime
}=a_{1}c_{2}+a_{2}c_{1},\text{ }\Delta _{3}^{\prime }=a_{1}c_{1}-a_{2}c_{2},
\notag \\
\Delta _{1}^{\prime \prime } &=&a_{3}+b_{3},\text{ }\Delta _{2}^{\prime
\prime }=a_{1}b_{2}+a_{2}b_{1},\text{ }\Delta _{3}^{\prime \prime
}=a_{1}b_{1}-a_{2}b_{2}.  \label{Deltas}
\end{eqnarray}

\subsection{Three-player Prisoners' Dilemma}

Prisoner's Dilemma (PD) is a noncooperative game \cite%
{Binmore,Rasmusen,Osborne} that is widely known in the areas of economics,
social, and political sciences. In recent years, quantum physics has been
added to this list. It was investigated early in the history of quantum
games and provided significant motivation for further work in this area.

Two-player PD is about two suspects, considered here as the players in a
game, who have been arrested on the allegations of having committed a crime
but there not not enough available evidence to convict them. The
investigators come up with an ingenious plan to make the suspects confess
their crime.

They are taken to separate cells and are not allowed to communicate. They
are contacted individually and, along with being dictated a set of rules,
are asked to choose between two choices (strategies): \emph{to Confess} $(%
\mathfrak{D})$ and \emph{Not to Confess} $(\mathfrak{C})$, where $\mathfrak{C%
}$ and $\mathfrak{D}$ stand for Cooperation and Defection. These are the
well-known wordings for the available choices for them and refer to the
choice they make to the fellow prisoner, and not to the authorities.

The rules state that if neither prisoner confesses, i.e. $(\mathfrak{C},%
\mathfrak{C})$, both are given freedom; when one prisoner confesses $(%
\mathfrak{D})$ and the other does not $(\mathfrak{C})$, i.e. $(\mathfrak{C},%
\mathfrak{D})$ or $(\mathfrak{D},\mathfrak{C})$, the prisoner who confesses $%
(\mathfrak{D})$ gets freedom as well as a financial reward, while the
prisoner who did not confess ends up in prison for a longer term. If both
prisoners confess, i.e. $(\mathfrak{D},\mathfrak{D})$, both are given a
reduced term.

In the two-player case, involving the players $A$ and $B$ the strategy pair $%
(\mathfrak{D},\mathfrak{D})$ comes out as the unique NE (and the rational
outcome) of the game, leading to the situation of both ending up in jail
with reduced term. The game offers a dilemma as the rational outcome $(%
\mathfrak{D},\mathfrak{D})$ differs from the outcome $(\mathfrak{C},%
\mathfrak{C})$, which is an available choice, and for which both prisoners
obtain freedom.

With the above notation, the three-player PD can be defined by making the
following associations%
\begin{equation}
S_{1}\sim \mathfrak{C},\text{ }S_{2}\sim \mathfrak{D},\text{ }S_{1}^{\prime
}\sim \mathfrak{C},\text{ }S_{2}^{\prime }\sim \mathfrak{D},\text{ }%
S_{1}^{\prime \prime }\sim \mathfrak{C},\text{ }S_{2}^{\prime \prime }\sim 
\mathfrak{D},
\end{equation}%
and afterwards imposing the following conditions \cite{3playerPD}:

a) The strategy $S_{2}$ is a dominant choice \cite{Rasmusen} for each
player. For Alice this requires%
\begin{equation}
\begin{array}{l}
\Pi _{A}(S_{2},S_{1}^{\prime },S_{1}^{\prime \prime })>\Pi
_{A}(S_{1},S_{1}^{\prime },S_{1}^{\prime \prime }), \\ 
\Pi _{A}(S_{2},S_{2}^{\prime },S_{2}^{\prime \prime })>\Pi
_{A}(S_{1},S_{2}^{\prime },S_{2}^{\prime \prime }), \\ 
\Pi _{A}(S_{2},S_{1}^{\prime },S_{2}^{\prime \prime })>\Pi
_{A}(S_{1},S_{1}^{\prime },S_{2}^{\prime \prime }),%
\end{array}%
\end{equation}%
and similar inequalities hold for players Bob and Chris.

b) A player is better off if more of his/her opponents choose to cooperate.
For Alice this requires%
\begin{equation}
\begin{array}{l}
\Pi _{A}(S_{2},S_{1}^{\prime },S_{1}^{\prime \prime })>\Pi
_{A}(S_{2},S_{1}^{\prime },S_{2}^{\prime \prime })>\Pi
_{A}(S_{2},S_{2}^{\prime },S_{2}^{\prime \prime }), \\ 
\Pi _{A}(S_{1},S_{1}^{\prime },S_{1}^{\prime \prime })>\Pi
_{A}(S_{1},S_{1}^{\prime },S_{2}^{\prime \prime })>\Pi
_{A}(S_{1},S_{2}^{\prime },S_{2}^{\prime \prime }),%
\end{array}%
\end{equation}%
and similar relations hold for Bob and Chris.

c) If one player's choice is fixed, the other two players are left in the
situation of a two-player PD. For Alice this requires%
\begin{equation}
\begin{array}{l}
\Pi _{A}(S_{1},S_{1}^{\prime },S_{2}^{\prime \prime })>\Pi
_{A}(S_{2},S_{2}^{\prime },S_{2}^{\prime \prime }), \\ 
\Pi _{A}(S_{1},S_{1}^{\prime },S_{1}^{\prime \prime })>\Pi
_{A}(S_{2},S_{1}^{\prime },S_{2}^{\prime \prime }), \\ 
\Pi _{A}(S_{1},S_{1}^{\prime },S_{2}^{\prime \prime })>(1/2)\left\{ \Pi
_{A}(S_{1},S_{2}^{\prime },S_{2}^{\prime \prime })+\Pi
_{A}(S_{2},S_{1}^{\prime },S_{2}^{\prime \prime })\right\} , \\ 
\Pi _{A}(S_{1},S_{1}^{\prime },S_{1}^{\prime \prime })>(1/2)\left\{ \Pi
_{A}(S_{1},S_{1}^{\prime },S_{2}^{\prime \prime })+\Pi
_{A}(S_{2},S_{1}^{\prime },S_{1}^{\prime \prime })\right\} ,%
\end{array}%
\end{equation}%
and similar relations hold for Bob and Chris.

Translating the above conditions while using the notation introduced in (\ref%
{symmetric_3player_game_definition}) requires%
\begin{equation}
\begin{array}{l}
\text{a) }\beta >\alpha ,\ \ \omega >\epsilon ,\ \ \theta >\delta , \\ 
\text{b) }\beta >\theta >\omega ,\ \ \alpha >\delta >\epsilon , \\ 
\text{c) }\delta >\omega ,\ \ \alpha >\theta ,\ \ \delta >(1/2)(\epsilon
+\theta ),\ \ \alpha >(1/2)(\delta +\beta ),%
\end{array}%
\end{equation}%
which defines the generalized three-player PD. For example \cite{3playerPD},
by letting

\begin{equation}
\alpha =7,\text{ }\beta =9,\text{ }\delta =3,\text{ }\epsilon =0,\ \omega =1,%
\text{ }\theta =5,  \label{PD_values}
\end{equation}%
all of these conditions hold.

\section{Three-player quantum Prisoners' Dilemma with GHZ state}

The values in (\ref{PD_values}) give $\gamma _{1}=-1$ and $\gamma _{2}=1$.
With the deltas given in (\ref{Deltas}), the Nash inequalities (\ref{NE_ABC}%
) take the form

\begin{gather}
\Pi _{A}(\vec{a}^{\ast },\vec{b}^{\ast },\vec{c}^{\ast })-\Pi _{A}(\vec{a},%
\vec{b}^{\ast },\vec{c}^{\ast })=  \notag  \label{NE_ABC_1} \\
\frac{1}{8}\left[ -(a_{3}^{\ast }-a_{3})(b_{3}+c_{3})+(a_{1}^{\ast
}-a_{1})(b_{1}c_{1}-b_{2}c_{2})-(a_{2}^{\ast }-a_{2})(b_{1}c_{2}+b_{2}c_{1})%
\right] \geq 0,  \notag \\
\Pi _{A}(\vec{a}^{\ast },\vec{b}^{\ast },\vec{c}^{\ast })-\Pi _{A}(\vec{a}%
^{\ast },\vec{b},\vec{c}^{\ast })=  \notag \\
\frac{1}{8}\left[ -(b_{3}^{\ast }-b_{3})(a_{3}+c_{3})-(b_{2}^{\ast
}-b_{2})(a_{1}c_{2}+a_{2}c_{1})+(b_{1}^{\ast }-b_{1})(a_{1}c_{1}-a_{2}c_{2})%
\right] \geq 0,  \notag \\
\Pi _{A}(\vec{a}^{\ast },\vec{b}^{\ast },\vec{c}^{\ast })-\Pi _{A}(\vec{a}%
^{\ast },\vec{b}^{\ast },\vec{c})=  \notag \\
\frac{1}{8}\left[ -(c_{3}^{\ast }-c_{3})(a_{3}+b_{3})-(c_{2}^{\ast
}-c_{2})(a_{1}b_{2}+a_{2}b_{1})+(c_{1}^{\ast }-c_{1})(a_{1}b_{1}-a_{2}b_{2})%
\right] \geq 0.  \notag \\
\end{gather}%
These inequalities show that for the PD game defined in (\ref{PD_values}),
no directional triplet can exist as a NE when the three players have the
choice to direct their respective unit vector along any direction i.e. there
are no restrictions placed on the players' directional choices.

The inequalities (\ref{NE_ABC}) suggest the following cases:

\subsection{Case (a)}

Consider $a_{3}=b_{3}=c_{3}=0$. Nash inequalities (\ref{NE_ABC}) then take
the form

\begin{eqnarray}
\Pi _{A}(\vec{a}^{\ast },\vec{b}^{\ast },\vec{c}^{\ast })-\Pi _{A}(\vec{a},%
\vec{b}^{\ast },\vec{c}^{\ast }) &=&\frac{1}{8}\gamma _{2}\left[
(a_{1}^{\ast }-a_{1})\Delta _{2}-(a_{2}^{\ast }-a_{2})\Delta _{3}\right]
\geq 0,  \notag \\
\Pi _{A}(\vec{a}^{\ast },\vec{b}^{\ast },\vec{c}^{\ast })-\Pi _{A}(\vec{a}%
^{\ast },\vec{b},\vec{c}^{\ast }) &=&\frac{1}{8}\gamma _{2}\left[
-(b_{2}^{\ast }-b_{2})\Delta _{2}^{\prime }+(b_{1}^{\ast }-b_{1})\Delta
_{3}^{\prime }\right] \geq 0,  \notag \\
\Pi _{A}(\vec{a}^{\ast },\vec{b}^{\ast },\vec{c}^{\ast })-\Pi _{A}(\vec{a}%
^{\ast },\vec{b}^{\ast },\vec{c}) &=&\frac{1}{8}\gamma _{2}\left[
-(c_{2}^{\ast }-c_{2})\Delta _{2}^{\prime \prime }+(c_{1}^{\ast
}-c_{1})\Delta _{3}^{\prime \prime }\right] \geq 0,  \notag \\
&&
\end{eqnarray}%
that can also be expressed as

\begin{eqnarray}
\Pi _{A}(\vec{a}^{\ast },\vec{b}^{\ast },\vec{c}^{\ast })-\Pi _{A}(\vec{a},%
\vec{b}^{\ast },\vec{c}^{\ast }) &=&\frac{1}{8}\gamma _{2}\left[ a_{1}^{\ast
}\Delta _{2}-a_{2}^{\ast }\Delta _{3}+\varsigma \right] \geq 0,  \notag \\
\Pi _{A}(\vec{a}^{\ast },\vec{b}^{\ast },\vec{c}^{\ast })-\Pi _{A}(\vec{a}%
^{\ast },\vec{b},\vec{c}^{\ast }) &=&\frac{1}{8}\gamma _{2}\left[
-b_{2}^{\ast }\Delta _{2}^{\prime }+b_{1}^{\ast }\Delta _{3}^{\prime
}+\varsigma \right] \geq 0,  \notag \\
\Pi _{A}(\vec{a}^{\ast },\vec{b}^{\ast },\vec{c}^{\ast })-\Pi _{A}(\vec{a}%
^{\ast },\vec{b}^{\ast },\vec{c}) &=&\frac{1}{8}\gamma _{2}\left[
-c_{2}^{\ast }\Delta _{2}^{\prime \prime }+c_{1}^{\ast }\Delta _{3}^{\prime
\prime }+\varsigma \right] \geq 0,
\end{eqnarray}%
where

\begin{equation}
\varsigma =a_{1}b_{2}c_{2}+a_{2}b_{1}c_{2}-a_{1}b_{1}c_{1}+a_{2}b_{2}c_{1}.
\end{equation}%
Consider the case when $\gamma _{2}>0,$ then for given $a_{1}^{\ast },$ $%
b_{1}^{\ast },$ and $c_{1}^{\ast }$, the restrictions on the directions that
the unit vectors $\vec{a},$ $\vec{b},$ and $\vec{c}$ can take can be
determined. For instance, for $a_{2}^{\ast },=b_{1}^{\ast },=c_{1}^{\ast }=1$%
, i.e. then $a_{2}^{\ast }=b_{2}^{\ast }=c_{2}^{\ast }=0,$these constraints
become

\begin{equation}
\Delta _{2}+\varsigma \geq 0,\text{ }\Delta _{3}^{\prime }+\varsigma \geq 0,%
\text{ }\Delta _{3}^{\prime \prime }+\varsigma \geq 0.
\end{equation}

\subsection{Case (b)}

Consider $\gamma _{2}=0$ and $a_{3}=b_{3}=c_{3}=0.$ With these constraints,
the allowed directions are confined to the X-Y plane and any directional
triplet then exists as a NE. In this case, from Eqs. (\ref{Deltas}) we then
have $\Delta _{1}=\Delta _{1}^{\prime }=\Delta _{1}^{\prime \prime }=0.$ As $%
\vec{a},$ $\vec{b},$ and $\vec{c}$ are unit vectors, we also have $a_{2}=\pm 
\sqrt{1-a_{1}^{2}},$ $b_{2}=\pm \sqrt{1-b_{1}^{2}},$ and $c_{2}=\pm \sqrt{%
1-c_{1}^{2}}$.

\section{Discussion}

We present an analysis of the three-partite EPR experiments that use a GHZ
state and is considered as a three-player non-cooperative quantum game. The
players' strategic choices are the three directions $\vec{a},$ $\vec{b},$
and $\vec{c}$ along which the dichotomic observables $\vec{n}.\vec{\sigma}$\
are measured, where $\vec{n}=\vec{a},\vec{b},\vec{c}$ and $\vec{\sigma}$ is
a vector whose components are the standard Pauli matrices. Using
Kaszlikowski and \.{Z}ukowski's results \cite{Kaszlikowski} for the quantum
probabilities involved in such experiments, we develop a three-player
quantum \ game, with the underlying three-partite EPR experiment. This
extends an approach to quantum games by Vaidman \cite{Vaidman} that does not
involve Hilbert space, and/or quantum measurement, and shows how
three-player quantum games with EPR experiments can be developed. Players'
strategies are their directions in terms of which their payoffs are
expressed using Eq. (\ref{QProbabilities}). Nash inequalities are used to
obtain Nash equilibria as direction triples and the players' payoffs are
then compared to their payoffs for the Nash equilibria in the classical game.

For a three-player Prisoners' Dilemma game, defined in (\ref{PD_values}), we
conclude that no directional triplet can exist as a NE when no restrictions
are placed on the players' directional choices. A directional triplet,
however, can exist as a NE under constraints placed on the directions
allowed to the players. This is in accordance with Eisert et al.'s\emph{\ }%
result in Ref. \cite{EWL} showing that a pair of unitary transformations $(%
\hat{Q},\hat{Q})$, where $\hat{Q}\sim \hat{U}(0,\pi /2),$ exists as a NE in
PD when the players' allowed actions are restricted to certain subsets of
the set SU(2) consisting of all unitary transformations.

As is known \cite{Benjamin1, FlitneyHollenberg} that the particular subset
of unitary transformations that Eisert et al. used in order to obtain the NE
of $(\hat{Q},\hat{Q})$ in two-player quantum Prisoners' Dilemma is not even
closed under composition.\emph{\ }In particular, in Eisert et al.'s protocol
for $2\times 2$ quantum games \cite{EWL}, the new Nash equilibria, and the
classical-quantum transitions that occur, are the outcomes of the particular
strategy space chosen that is a two-parameter subset of single qubit unitary
operators. By choosing a different, but equally plausible, two-parameter
strategy a different Nash equilibria with different classical-quantum
transitions can arise.

Using an EPR setting, and a shared GHZ state, for a three-player quantum
Prisoners' Dilemma game, we present an approach that is driven along purely
probabilistic lines with only an implicit reference to the mathematical
formalism of quantum theory, and showing the constraints on the players'
directional choices under which a particular triplet can exist as a NE in
the game.

\end{document}